\documentclass[a4paper]{article}
\usepackage{RR}
\RRNo{6528}
\usepackage{hyperref}
\RRdate{May 2008}

\usepackage{graphicx}  
\usepackage{url}       
\usepackage{amsmath,amsfonts,amssymb}   
\usepackage{color}
\usepackage{theorem}

\newcommand{\nnet}{\net}

\newcommand{\placeset}{\mathcal{P}}
\newcommand{\transset}{\mathcal{T}}
\newcommand{\flowset}{\mathcal{F}}
\newcommand{\datefamilies}{\mathbb{T}}
\newcommand{\dateinit}{\mathbb{T}_{\rm init}}

\newcommand{\ie}{\emph{i.e}}
\newcommand{\eg}{\emph{e.g}}
\newcommand{\preset}[1]{{{}^\bullet{#1}}}
\newcommand{\postset}[1]{{#1}^\bullet}

\newcommand{\cone}[1]{\lceil#1\rceil}

\newcommand{\config}{\kappa}
\newcommand{\cluster}{\mathbf{c}}
\newcommand{\maxconfig}{\overline{\kappa}}

\newcommand{\cocnets}{\emph{CO}-nets }

\newcommand{\Tfam}{\Delta }

\newcommand{\beqq}{\begin{eqnarray*}}
\newcommand{\eeqq}{\end{eqnarray*}}
\newcommand{\beq}{\begin{eqnarray}}
\newcommand{\eeq}{\end{eqnarray}}
\newcommand{\bea}{\begin{array}}
\newcommand{\eea}{\end{array}}
\newcommand{\proof}{\paragraph{Proof:}}
\newcommand{\eproof}{\hfill$\diamond$}

\newcommand{\daemon}{\omega}
\newcommand{\Daemons}{\Omega}
\newcommand{\orchnet}{OrchNet}
\newcommand{\preorchnet}{pre-OrchNet}
\newcommand{\wfnet}{WFnet}
\newcommand{\prewfnet}{pre-WFnet}
\newcommand{\Orchnet}{\mathcal{N}}
\newcommand{\Preorchnet}{\mathbb{N}}
\newcommand{\WFnet}{\mathcal{W}}
\newcommand{\PreWFnet}{\mathbb{W}}
\newcommand{\net}{N}
\newcommand{\wnet}{W}
\newcommand{\Tinit}{T_{{\rm init}}}

\newcommand{\concurrent}{\|}
\newcommand{\conflict}{\#}

\newcommand{\MaxConfigs}[1]{\overline{\mathcal{V}}\left(#1\right)}

\newtheorem{definition}{Definition}
\newtheorem{theorem}{Theorem}

\newtheorem{assumption}{Assumption}

\RRauthor{Anne Bouillard
  \and
Sidney Rosario
 \and
Albert Benveniste
\and
Stefan Haar
\thanks{
S.Rosario and A. Benveniste are with Irisa/Inria, Campus de Beaulieu,
Rennes. A. Bouillard is with Irisa/ENS Cachan Campus de Ker Lann
and S.Haar is with Alcatel-Lucent Bell Labs, Kanata, ON,Canada.
}
}

\RRtitle{Monotonie dans les orchestrations de web services }
\RRetitle{Monotony in Service Orchestrations
\thanks{
This work was partially funded by the ANR national research program
DOTS (ANR-06-SETI-003), DocFlow (ANR-06-MDCA-005) and the project
CREATE ActivDoc.
} }

\titlehead{Monotony in Service Orchestrations}

\RRresume{
Les orchestrations de services web sont des compositions de services
\'el\'ementaires.  Ces services, fournissent un 'contrat' \`a l'orchestrateur,
ce qui garantit une certaine performance de leur service.  Ces
contrats sont utilis\'es par l'orchestrateur pour proposer un contrat
\`a un client pour son propre service.  Cela se fait par la 'compostion de
contrats'. Du point vue de la performance, la composition de contrats suppose implicitement que
"L'am\'elioration de la performance d'un service va rendre
l'orchestration plus performante". La composition de contrats suppose ainsi que
les orchestrations sont "monotones".

Dans quelques orchestrations, cependant, la monotonie peut ne pas \^etre
respect\'ee. Lorsque la performance d'un service s'am\'eliore, la performance
de l'orchestration se d\'egrade. Ceci est tr\`es g\^enant car cela rend le
processus de composition de contrats invalide.
Dans ce rapport, nous d\'efinissons la monotonie pour les orchestrations
mod\'elis\'ees par des r\'eseaux d'occurrence color\'es (CO-nets) et nous
caract\'erisons  la classe des orchestrations monotones. Nous d\'emontrons que
tr\`es peu d'orchestrations sont monotone en pratique, ce qui est largement d\^u \`a la
possibilit\'e d'am\'eliorer la latence en d\'egradant la qualit\'e de la
r\'eponse donn\'e. Nous proposons ensuite un raffinement de la monotonie,
la "monotonie conditionnelle", qui interdit ce type de 'triche'. Nous
montrons que la monotonie conditionnelle est tr\`es g\'en\'eralement satisfaite par
les orchestrations. Cette \'etude nous m\`ene \`a reconsid\'erer la
formulation des contrats dans le cadre des orchestrations de services web.
}

\RRabstract{
Web Service orchestrations are compositions of different Web Services
to form a new service. The services called during the orchestration
guarantee a given performance to the orchestrater, usually in the form
of contracts. These contracts can be used by the orchestrater to
deduce the contract it can offer to its own clients, by performing
contract composition. An implicit assumption in contract based QoS
management is: "the better the component services perform, the better
the orchestration's performance will be". Thus, contract based QoS
management for Web services orchestrations implicitly assumes
monotony.

In some orchestrations, however, monotony can be violated, i.e., the
performance of the orchestration improves when the performance of a
component service degrades. This is highly undesirable since it can
render the process of contract composition inconsistent.

In this paper we define monotony for orchestrations modelled by
Colored Occurrence Nets (CO-nets) and we characterize the classes of
monotonic orchestrations. We show that few orchestrations are indeed
monotonic, mostly since latency can be traded for quality of
data. We also propose a sound refinement of monotony, called
\emph{conditional monotony}, which forbids this kind of cheating and
show that conditional monotony is widely satisfied by orchestrations.
This finding leads to reconsidering the way SLAs should be formulated.
}

\RRmotcle{web service, orchestrations, contrats, monotonie}
\RRkeyword{web service, orchestrations, contracts, monotony}

\RRprojets{Distribcom}
\RRtheme{\THCom}

\begin{document}

\RCRennes
\makeRR

\tableofcontents

\section{Introduction}

Web Services and their compositions are being widely used to build
distributed applications over the internet. Web Service orchestrations 
are compositions of Web Services to form an aggregate, and usually
more complex, Web Service (WS). The services involved in a WS orchestration
may have widely different behaviour (both functional and non-functional) 
and may be separated geographically by large distances. 

A WS orchestration is itself a Web Service, and it can be further composed 
with other Web Services to build increasingly sophisticated services. The 
functioning of such service compositions can thus be quite complex in nature.
There is a need to formally describe these systems in order to be able to build 
and reason about them. Different 
formalisms have been proposed for this purpose, the most popular
amongst these is the Business Process Execution Language
(BPEL)~\cite{bpel4wsspec} a standard proposed by Microsoft and IBM.
Another formalism is Orc~\cite{orc} a small and elegant formalism
equipped with extensive semantics work~\cite{orctrace,orc-wsfm07}.
Various models exist, that have been either used to model directly orchestrations or
choreographies, or as a semantic domain for some formalisms. Noticeably
Petri Nets based models, e.g., WorkFlow Nets~\cite{aalst98application,Aalst97} and process algebra based models. 

 The main focus of the existing models is to capture the functional aspects of 
such compositions. 
However, non-functional --- also called Quality of Service (QoS) ---
aspects involved in services and their compositions need also to be
considered.  The QoS of a service is characterised by different
metrics, for \textit{e.g.,} latency, availability, throughput,
security, etc. Standard WSLA~\cite{WSLA} specifies how QoS can be
specified using Service Level Specifications and Agreements. Examples
of SLA parameters are found in this document that illustrate the
current practice. In particular, ``conditional SLA obligations''
consist in specifying what a service must guarantee given that the
environment satisfies certain assumptions.

QoS management is typically based on the notion of \emph{contract}.
Contracts are agreements made between the orchestrater and the
different actors (the called services) involved in the
orchestration. Contracts formalise the duties and responsibilities
each subcontractor must satisfy. For \emph{e.g.,} a service which is
called in an orchestration can have a contract of the type : \emph{for
95\% of the requests the response time will be less than 5ms}.  All
these contracts with the services involved in the orchestration could
then be composed by the orchestrater to help it propose its own
contract with users of the orchestration.  This process is called
\emph{contract composition.}

In~\cite{QoS-icws07} we introduced the notion of \emph{probabilistic
contracts} to formalise the QoS behaviour of services --- the work of~\cite{QoS-icws07} focused on latency. We showed how
these contracts can be composed to get the end-to-end orchestration's
contract. We also showed that realistic \emph{overbooking} of
resources by the orchestrater is possible using this approach.

Contract based QoS management relies on the implicit assumption that,
if each sub-contractor meets its contract objectives, then so does the
orchestrater. Vice-versa, a sub-contractor breaching its contract can
cause the orchestrater to fail meeting its overall contract
objectives. Thus the whole philosophy behind contracts is that the
better sub-contractors behave, the better the overall orchestration
will meet its own contract. In fact, the authors themselves have
developed their past work~\cite{QoS-icws07} based on this credo\dots until
they discovered that this implicit assumption could easily be
falsified. Why so? 

\begin{figure}[h]
\centerline{\input{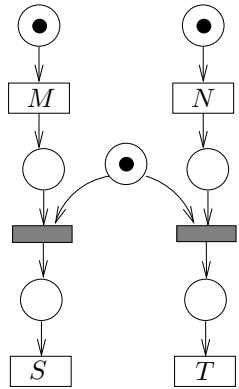}}
\caption{A non-monotonic orchestration}
\label{fig::non-monotone-example}
\end{figure}
As an illustration example, consider the orchestration in
Figure~\ref{fig::non-monotone-example}. Services $M$ and $N$ are
first called in parallel. If $M$ responds first, service $S$ is
next called and the response of $N$ is ignored. If $N$ responds
first, $T$ is called and not $S$. Let $\delta_i$ denote the
response time of site $i$. Assume the following delay behaviour :
$\delta_M < \delta_N$ and $\delta_S \gg \delta_T$. Since $M$ responds
faster, the end-to-end orchestration delay $d_0 = \delta_M+\delta_S$.
Now let service $M$ behaves slightly 'badly', \ie\ delay $\delta_M$ increases
and becomes slightly greater than $\delta_N$. Now service $T$ is called and
the new orchestration delay is $d_1=\delta_N+\delta_T$. But since
$\delta_S \gg \delta_T$, $d_1$ is in fact lesser than $d_0$.
This orchestration is non-monotonic since increasing the
latency of one of its components can decrease the end-to-end latency
of the orchestration. So, what is the nature of the difficulty?

``Simple'' composed Web services are such that QoS aspects do not
interfere with functional aspects and do not interfere with each
other. Their flow of control is typically rigid and does not involve
if-then-else branches. For such cases, latencies will compose gently
and will not cause pathologies as shown above. However, as evidenced
by the rich constructions offered by BPEL, orchestrations and
choreographies can have branching based on data and QoS values,
various kinds of exceptions, and timers. With such flexibility, 
non-monotony such as that exhibited by the example of 
Figure~\ref{fig::non-monotone-example} can very easily occur. 

\emph{Lack of monotony, in turn, impairs using
contracts for the compositional management of QoS.} Surprisingly
enough, this fact does not seem to have been noticed in the literature.

In this paper we give a classification for orchestrations based on their
monotonic characteristics.  Our study focuses on latency, 
although other aspects of QoS are discussed as well.
This paper is organised as follows: Section~\ref{sec:erpifuerhfu} informally
introduces the notion of monotony with examples. In section \ref{sec:orchnet} we
recall the definition of Petri nets and colored Petri nets, and
introduce our model, \orchnet. 
A formal definition of monotony and a 
characterisation of monotonic orchestrations is then given in 
section~\ref{sec:char-monotony}. Section \ref{sec:refined-qos}
introduces \emph{conditional} {monotony} which will be useful to deal
with non-monotonic orchestrations which use data-dependent control.

\section{Non-monotonic patterns in {\cocnets}}
\label {sec:erpifuerhfu}
We now show examples of non-monotonic orchestrations and identify the
source of their non-monotony. The necessary and sufficient conditions
for monotony that we give later were inspired from the study of these
patterns.  The examples we discuss here informally use Petri nets; we
believe they are self-explanatory and do not deserve formal
definitions. The formal definitions of Petri nets and their semantics
is given in section~\ref{sec:orchnet}.

\begin{figure}[h]
  \centering
  \input{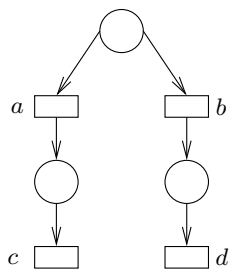}
  \caption{Choices are not monotonic.}
  \label{fig:choices}
\end{figure}

Consider the net on Figure~\ref{fig:choices} with four transitions
$a,b,c$ and $d$ with latencies $\tau_a,\tau_b,\tau_c$ and 
$\tau_d$ respectively. Let $\tau_{a} = 2$,
$\tau_{b} = 3$, $\tau_{c} = 4$ and $\tau_{d} =7$. Here $a$ fires
before $b$ and the overall orchestration latency is $\tau_a+\tau_c=6$.
Now, if $\tau_{b}=1$, $b$ fires before $a$ and so the orchestration latency
is $\tau_b+\tau_d=8$. Since decreasing the latency of $b$ increases the
orchestration's latency, the net is not monotonic.  
Non-monotony arises from the fact that the futures of the conflicting
transitions $a$ and $b$ had different latencies. The net is monotonic if
the latencies of $c$ and $d$ are constrained to be the same.

\begin{figure}[h]
  \centering
  \input{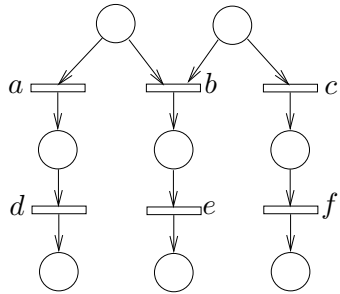}
  \caption{Choice and Concurrency inter-playing}
  \label{fig::non-monotone-eg2}
\end{figure}

In Figure~\ref{fig::non-monotone-eg2}, transitions $a$ and $c$ are concurrent,
$a\conflict b$ and $b\conflict c$. Since we saw in the previous example that
the futures of conflicting transitions should have the same latencies, let
us set $\tau_d=\tau_e=\tau_f=\tau_*$. If $\tau_a=4,\tau_b=3,\tau_c=5$, $b$ fires
first and the overall latency is $\tau_b+\tau_e=4+\tau_*$. Now if $\tau_a=2$, $a$
fires first, $b$ is blocked and $c$ will eventually fire. The overall latency 
here is $max(\tau_a+\tau_d,\tau_c+\tau_f)=5+\tau_*$.
We thus see that for this net to be monotonic, we must have the latencies of
the concurrent transitions $a$ and $c$ to be the same (along with their futures).
In fact, since $\tau_a=\tau_c$ and $\tau_d=\tau_f$, the two concurrent branches
can be "folded" into a single branch to give us a net similar to that of 
Figure~\ref{fig:choices}.

~\\\noindent{\emph{A monotonic orchestration:}
Figure~\ref{fig:monotone-eg} shows an orchestration that has a
fork-join pattern (in the left). The right side of the figure 
shows the net's \emph{unfolding}. 
(The unfolding of a net $N$ is an acyclic net which includes all the
possible executions of $N$. Every place in the unfolding can be marked by
at-most one transition and so every possible way of enabling a transition is
distinguished. In Figure~\ref{fig:monotone-eg}, the two possible cases in 
which event $c$ can be fired are distinguished in the unfolded net).
Since the left net has a choice pattern, one
may think that it is not monotonic. However, this is not true. 
From the previous example we know that the unfolded net is monotonic
since the futures of the conflicting transitions $a$ and $b$ are the
same (and hence have the same latencies).

\begin{figure}[h]
  \centering
  \input{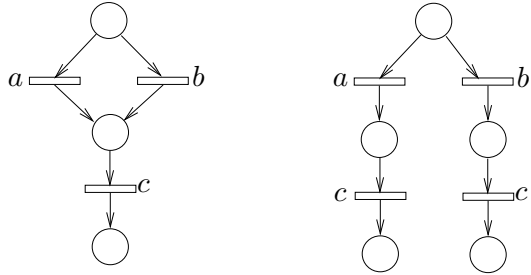}
  \caption{A monotonic orchestration with a fork-join pattern (left)
  and its unfolding (right).}
  \label{fig:monotone-eg}
\end{figure}

At this point we have identified good and bad patterns for
monotony. Going beyond such simple findings proved to be surprisingly
challenging. In particular, characterizing that the restricted class
of orchestrations we define is a \emph{necessity} for monotony is
demanding. All this requires more formal material that we introduce next.

\section{The Orchestration Model:\ \orchnet s}
\label{sec:orchnet}

 In this section we present the orchestration model that we use
for our studies, which we call \emph{\orchnet s}. \orchnet s
are a special form of \emph{colored occurrence
nets (CO-nets)}, which are high level Petri Nets.

We have chosen this mathematical model for the following reasons. From
the semantic studies performed for BPEL~\cite{bpmBpelAnalysisReport05,Fisteus2005} and
Orc~\cite{orctrace,orc-wsfm07} we need to support in an elegant and succinct way
the following features: concurrency, rich control patterns including
preemption, and for some cases even recursion. The first two
requirements suggest using colored Petri nets. The last requirement
suggests considering extensions of Petri nets with
dynamicity. However, in our study we will not be interested in the
specification of orchestrations, but rather in their
executions. Occurrence nets are concurrent models of executions of
Petri nets. As such, they encompass orchestrations involving recursion
at no additional cost.
The executions of Workflow Nets~\cite{aalst98application,Aalst97} are also (CO-nets).

\subsection{Petri nets, Occurrence nets}
\label {perofuiheriofuh}
\begin{definition} A \emph{Petri net} is a tuple $\nnet=(\placeset,\transset,\flowset,M_0)$,
where
\begin{itemize}
\item $\placeset$ is a set of \emph{places},
\item $\transset$ is a set of \emph{transitions} such that $\placeset\cap\transset=\emptyset$,
\item $\flowset\subseteq(\placeset\times\transset)\cup(\transset\times\placeset)$ is a set of \emph{flow arcs},
\item $M_0:\placeset\to\mathbf{N}$ is the \emph{initial marking}.
\end{itemize}
\end{definition}
The elements in $\placeset\cup\transset$ are called the \emph{nodes} of $\nnet$
and will be denoted by variables for \eg, $x$.
For node $x\in\placeset\cup\transset$, we call $\preset{x}=\{y\ |\ (y,x)\in\flowset\}$
the \emph{preset} of $x$, and $\postset{x}=\{y\ |\ (x,y)\in\flowset\}$ the \emph{postset} of $x$. 
A \emph{marking} of the net is a multiset $M$ of places, \ie\ a map from $\placeset$ to $\mathbf{N}$.
A transition $t$ is \emph{enabled} in marking $M$ if $\forall p\in\preset{t}, M(p) > 0$.
This enabled transition can \emph{fire} resulting in a new marking $M-\preset{t}+\postset{t}$
denoted by $M[t\rangle M'$. A marking $M$ is \emph{reachable} if there exists a sequence
of transitions $t_0,t_1\ldots t_n$ such that $M_0[t_0\rangle M_1[t_1\rangle \ldots [t_n\rangle M$.
A net is \emph{safe} if for all reachable markings $M$,
$M(p) \subseteq \{0,1\}$ for all $p\in\placeset$.
%
We define two relations on the nodes of a net: 

\begin{definition}[Causality]
\label{defn::causality}
For a net $\nnet=(\placeset,\transset,\flowset,M_0)$ the \emph{causality relation}
$<$ is the transitive closure of the relation $\prec$ defined as:
\begin{itemize}
\item If $p\in\preset{t}$, then $p\prec t$,
\item If $p\in\postset{t}$, then $t\prec p$,
\end{itemize}
where $p\in\placeset,t\in\transset$.
\end{definition}
The reflexive closure of $<$ is denoted by $\leq$.
For a node $x\in\placeset\cup\transset$, the set of \emph{causes}
of $x$ is $\cone{x}=\{y\in\placeset\cup\transset\ |\ y\leq x\}$.

\begin{definition}[Conflict]
\label{defn::conf}
For a net $\nnet=(\placeset,\transset,\flowset,M_0)$ two nodes
$x$ and $y$ are in \emph{conflict} - denoted by $x\conflict y$ - if there 
exist distinct transitions $t,t' \in T$,
such that $t \leq x, t'\leq y$ and $\preset{t}\cap\preset{t'}\ne\emptyset$.
\end{definition}
Nodes $x$ and $y$ are said to be \emph{concurrent} - written as $x\concurrent y$ -
if neither $(x\leq y)$ nor $(y\leq x)$ nor $(x\conflict y)$.
A set of concurrent conditions $P\subseteq \placeset$ is called a \emph{co-set}.
A \emph{cut} is a maximal (for set inclusion) co-set. 

\begin{definition}[Configuration]
A \emph{configuration} of $\nnet$ is
a subnet $\config$ of nodes of $\nnet$ such that:
\begin{enumerate}
\item $\config$ is \emph{causally closed}, \ie, if $x < x'$ and $x'\in\config$ then $x\in\config$
\item $\config$ is \emph{conflict-free}, \ie, for all nodes $x,x' \in\config, \lnot(x\conflict x')$
\end{enumerate}
\end{definition}
For convenience, we will assume that the maximal nodes (w.r.t the $<$ relation)
in a configuration are places.

\begin{definition}[Occurrence nets]
\label{defn::onet}
A safe net $\nnet=(\placeset,\transset,\flowset,M_0)$ is
called an \emph{occurrence net ({O-net})} iff
\begin{enumerate}
\item\label{defn::onet-cond1} $\lnot(x\conflict x)$ for every $x\in \placeset\cup\transset$.
\item\label{defn::onet-cond2} $\leq$ is a partial order and $\cone{t}$ is finite
      for any $t\in\transset$.
\item\label{defn::onet-cond3} For each place $p\in\placeset$, $|\preset{p}| \leq 1$.
\item\label{defn::onet-cond4} $M_0 = \{p\in\placeset|\preset{p}=\emptyset\}$, \ie\ the initial marking
      is the set of minimal places with respect to $\leq_\nnet$.
\end{enumerate}
\end{definition}
%
Occurrence nets are a good model for representing the possible
executions of a concurrent system.  \emph{Unfoldings} of a
safe Petri net, which collect all the possible executions of the net,
are occurrence nets. Unfoldings are defined as follows.

For $\net$ and $\net'$ two safe
nets, a map $\varphi:\placeset\cup\transset\mapsto
\placeset'\cup\transset'$ is called a \emph{morphism} of $\net$ to
$\net'$ if: 1/ $\varphi(\placeset)\subseteq\placeset'$ and 
$\varphi(\transset)\subseteq\transset'$, and 2/ for every $t\in\transset$
and $t'=\varphi(t)\in\transset'$,
$\preset{t}\cup\{t\}\cup\postset{t}$ is in bijection with 
$\preset{t'}\cup\{t'\}\cup\postset{t'}$ through $\varphi$.

Now, for $\net$ a safe net, there exists pairs $(U,\varphi)$ where $U$
is an occurrence net and $\varphi: U\mapsto\net$ is a morphism --- the
places and transitions of $U$ are ``labeled'' with places and
transitions of $\net$ through morphism $\varphi$. The \emph{unfolding}
of $\net$, denoted by $(U_\net,\varphi_\net)$ or $U_\net$ for short,
is the smallest pair $(U,\varphi)$ with the above properties, where
smallest refers to inclusion up to isomorphism. The configurations of
$U_\net$ are the executions of $\net$, seen as partial orders of
events.

\begin{figure}[h]
\centerline{\input{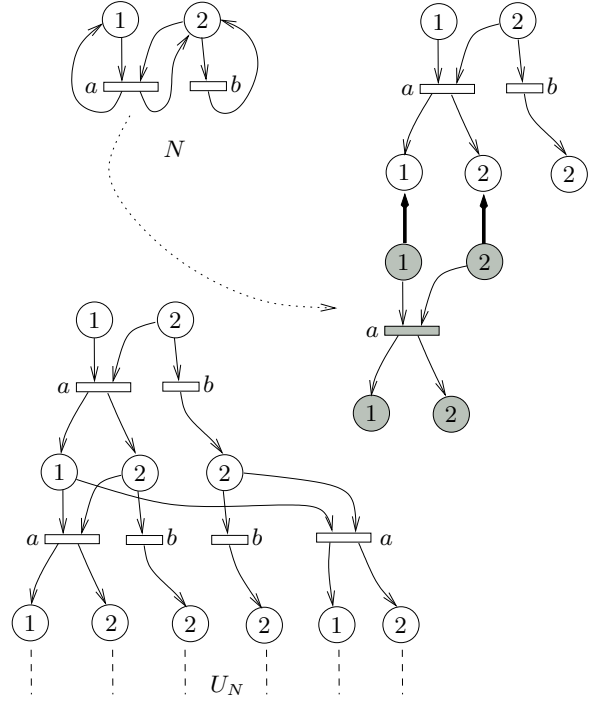}}
\caption{Construction of the unfolding $U_N$ of a safe net $N$.}
\label{fig::unfolding}
\end{figure}
Figure~\ref{fig::unfolding} shows a safe net $N$ and a small part
of its unfolding $U_N$. A step in the construction of $U_N$ is shown 
in the figure on the right. In each such step, a set of concurrent 
conditions $X$ in $U_N$ are chosen such that $\varphi(X)=\preset{t}$ for
some transition $t$ in $N$. The set of nodes 
$X\cup\{t'\}\cup\postset{t'}$ are then added to $U_N$ where
$\varphi(t')=t$ and $\varphi(\postset{t'})=\postset{t}$.
The right figure shows the addition of a new copy of transition $a$ (along with
its preset and postset) to $U_N$.

Any place of an occurrence net (specifically, an unfolding) gets
a token \emph{at-most} once. We can thus talk about "the token of the place"
unambiguously for any place in these nets.

\subsection{Our Model: \orchnet s}
\label{sec::our-model}

  We now present the orchestration model that we use
for our studies, which we call \emph{\orchnet s}. \orchnet s
are occurrence nets in which tokens are equipped with attributes
(or colors). 
%
%
\begin{figure}[h]
\centerline{\input{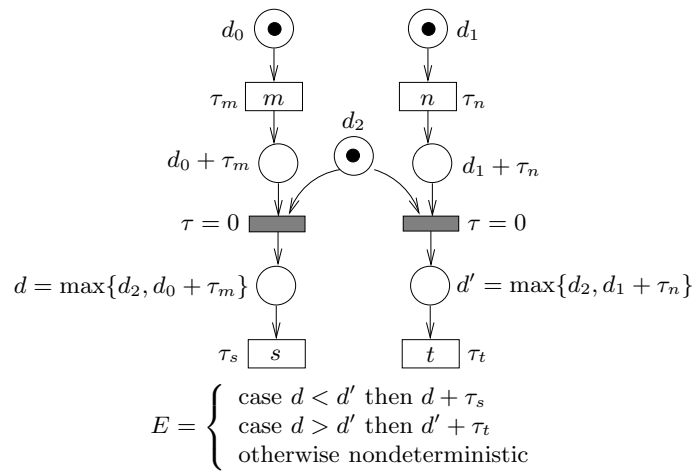}}
\caption{A Generic Transition of our Model. The arc
expressions are shown next to the arcs and the guard
expression is written next to the transition.}
\label{fig::generic-coc-transition}
\end{figure}
Figure~\ref{fig::generic-coc-transition} shows an \orchnet\ equipped with attributes related to timing.
Places are labeled with {dates} which is in fact the date of the token of that place.
Transitions are labeled with {latencies}. The tokens in the three minimal
places are given \emph{initial dates} (here, $d_0, d_1, d_2$). 
The four named transitions $m,n,s$ and $t$ are labeled with \emph{latencies} 
$\tau_m,\tau_n,\tau_s$ and $\tau_t$ respectively, and the two shaded transitions have zero latency.

The presence of dates in tokens alters the firing semantics slightly.
A transition $t$ is enabled at date when all places in its preset have tokens. 
It then takes $\tau_t$ additional time to fire. For example, the
shaded transition in the left has all its input tokens at 
$\max\{d_2,d_0+\tau_m\}$ and so it fires at $\max\{d_2,d_0+\tau_m\}+0$ 
since it has zero latency. If a transition fires at date $d$, then 
the tokens in its postset have the date $d$. This is shown in the figure, e.g., on
the place following the left shaded transition, which has date $\max\{d_2,d_0+\tau_m\}$.

When  transitions are in conflict, (e.g., the two shaded transitions 
in Figure~\ref{fig::generic-coc-transition}), the transition that
actually occurs is governed by a \emph{race policy}~\cite{MarsanBCDF98}.
If a set of enabled transitions are in conflict, 
the one with smallest date of occurrence will fire, 
preempting the other transitions in conflict with it. So,
in Figure~\ref{fig::generic-coc-transition}, the left shaded transition
or the right shaded transition will fire depending on whether $d<d'$ or
$d>d'$ respectively, with a nondeterministic choice if $d=d'$. 
This results in selecting the left most or right most continuation 
(firing $s$ or $t$) depending on the above cases.
The resulting overall latency $E$ of the orchestration is
shown at the bottom of the figure.


In addition to dates, tokens in \orchnet s can have {data attributes}, which we call \emph{values}.
We have not shown this in Figure~\ref{fig::generic-coc-transition}, in order to keep it simple.
Values of tokens in the preset of a transition $t$ can be combined by a 
value function $\phi_t$ attached to $t$. The resulting value is taken by the token in the postset of $t$.

Transitions can also have \emph{guards} (the presence of such a guard is not necessary). 
Guards are boolean predicates involving values of the tokens 
in the transition's preset. A transition is enabled only if its guard is true. 
An ``If(cond)-then-else'' branching can be implemented by using the following mechanism: 
Put a place $p$, followed by two transitions $t, t'$. Guard $t$ with the predicate
``cond=true'' and $t'$ with the predicate ``cond=false''.

At this point we are ready to provide the formal definition of \emph{\orchnet s}:

\begin{definition} [\orchnet] \label {perofuihpeuioh}
An \emph{\orchnet} is a tuple $\Orchnet=(\net,\Phi,T,\Tinit)$ consisting of
\begin{itemize}
\item An occurrence net $\net$ with token attributes 
$c=(value,date)$.
\item A family $\Phi=(\phi_t)_{t\in\transset}$ of \emph{value functions,}
whose input parameters are the values of the transition's input tokens.
\item A family $T=(\tau_t)_{t\in\transset}$ of \emph{latency functions,}
whose input parameters are the values of the transition's input tokens.
The range of the latency functions is in $\mathbf{R}\cup\{+\infty\}$
\item A family $\Tinit=(\tau_p)_{p\in\min(\placeset)}$ of \emph{initial date
functions} for the minimal places of $\net$.
The range of the initial date functions is in $\mathbf{R}\cup\{+\infty\}$
\end{itemize}
\end{definition}
We do not specify a concrete range of value functions since they can be any
arbitrary value and this is not of direct importance in the sequel. How a
transition $t$ modifies the attributes of tokens is formalized now: Let the
preset of $t$ have $n$ places whose tokens have $(value,date)$ attributes
$(v_1,d_1)\ldots (v_n,d_n)$.  Then all the tokens in the postset of $t$ have
the pair $(v_t,d_t)$ of value and date, where:
\beq v_t &=& \phi_t(v_1\ldots v_n) \nonumber \\ d_t &=& \max\{d_1\ldots
d_n\}+\tau_t(v_1\ldots v_n) \label{eqn:2rmn2t2} \eeq
The \emph{race policy} during execution is formalized as follows: In any given
marking $M$, let $T$ be the set of transitions that are \emph{possibly
enabled}, i.e. $\forall t\in T$, $\preset{t}$ is marked in $M$.  Then the
transition $t$ that is \emph{actually enabled}, (which really fires) is given
by: \[ t = \textrm{arg}\,\underset{t \in T}{\textrm{min\ }} d_t \] where
$\textrm{arg}\,\underset{x \in X}{\textrm{min\ }f(x)}=x^*\in X$ s.t. $\forall
x'\in X,f(x^*) \leq f(x')$.

If two transitions have the same $d_t$, then the choice of the transition that actually fires is non-deterministic.
If for a transition $t$, the delay $d_t$ is infinite, it is equivalent to saying that transition $t$ does not occur.

%
%

The choice extensions to occurrence nets in \orchnet s is inspired by the
application domain: compositions of web services. It reflects the following facts. 
\begin{itemize}
	\item Since we focus on latency, $(value,date)$ is the only needed color.
	\item Orchestrations rarely involve decisions on actions based on absolute dates. Timeouts are an exception, but these can be 
modelled explicitly, without using dates in guards of transitions. 
        This justifies the fact that guards only have token values as inputs, and not their dates.
  \item The time needed to perform transitions does not depend on the tuple of dates $(d_1\ldots d_n)$ 
        when input tokens were created, but it can depend on the data $(v_1\ldots v_n)$ and computation 
        $\phi$ performed on these. This justifies our restriction for output arc expressions.
\end{itemize}

If it is still wished that control explicitly depends on dates, then dates must be measured and can then be stored as part of the value $v$.

\paragraph*{Actually Occurring Configuration and Execution Time}
In general, both value and latency functions can be nondeterministic. We
introduce an invisible daemon variable $\daemon$ that resolves this
nondeterminism and we denote by $\Daemons$ its domain. For a given value 
of $\daemon$, the value and latency functions $\phi_t^{\daemon}$ and $\tau_t^{\daemon}$
are deterministic functions.

Let $\Orchnet=(\net,\Phi,T,\Tinit)$ be a finite \orchnet.
For a value $\daemon\in\Daemons$ for the daemon we can calculate the 
following \emph{dates} for every transition $t$ and place $p$ of $\Orchnet$:
\beq
\bea{rcl}\displaystyle
d_p(\daemon) &=& \left\{\bea{ll}
\tau^{\daemon}_p & \mbox{if $p$ is minimal}
\\
d_s(\daemon) & \mbox{where $s=\preset{p}$ otherwise}
\eea\right.
\\
d_t(\daemon) &=& \max\{d_n(\daemon)\mid n\in\preset{{t}}\}+
\tau_t^{\daemon}(v_1,\ldots v_n)
\eea
\label {pf349fuyeuir}
\eeq
where $v_1,\ldots v_n$ are the value components of the tokens 
in $\preset{t}$ as in equation~(\ref{eqn:2rmn2t2}).
If $\config$ is a configuration of $\net$, the future
$\Orchnet^\config$ is the \orchnet\ $(\net^\config,\Phi_{\net^\config},T_{\net^\config},\Tinit')$
where $\Phi_{\net^\config}$ and $T_{\net^\config}$ are the restrictions of
$\Phi$ and $T$ respectively, to the transitions of $\net^\config$.
$\Tinit'$ is the family derived from $\Orchnet$ according to (\ref{pf349fuyeuir}): for any minimal
place $p$ of $\net^\config$, the initialisation function is given by ${\tau'_p}^{\daemon} = d_p(\daemon)$.
For $X$ a set of nodes of $\net$, let
\beqq
\transset_{\min}(X) &=& \{
t\in\transset(X)\mid \preset{\preset{t}}\cap X = \emptyset
\}
\eeqq
Now define inductively,
\beq
\config_{0}(\daemon) &=& \emptyset
\nonumber \\
\config_m(\daemon) &=& \config_{m-1}(\daemon)\cup\{t_m\}\cup\preset{t_m}\cup\postset{t_m}
\label {eroifuego}
\\
&& \mbox{where } t_m =
\arg\min_{t\in\transset_{\min}(\net^{\config_{m-1}(\daemon)})} d_t(\daemon)
\nonumber
\eeq
Since net $\net$ is finite, the above inductive definition terminates
in finitely many steps when $ \net^{\config_{m}(\daemon)}=\emptyset$.
Let $M(\daemon)$ be this number of steps. We thus have 
\[
\emptyset=\config_0\subset\config_1(\daemon)\dots
\subset\config_{M(\daemon)}(\daemon)
\]
$\config_{M(\daemon)}(\daemon)$ is a maximal configuration which actually occurs 
according to our timed semantics, for a fixed $\daemon$.
Each step decreases the number of maximal configurations sharing
$\config_m(\daemon)$ as a prefix. 
%
We denote by $\maxconfig(\Orchnet,\daemon)$, the
maximal configuration $\config_{M(\daemon)}(\daemon)$ 
that actually occurs.

For a prefix $B$ of $\net$ define
\beq
E_\daemon(B,\Orchnet) = \max \{d_x(\daemon) \mid x\in B\} 
\label {p;fuhu}
\eeq
If $B$ is a configuration, then 
$E_\daemon(B,\Orchnet)$ is the time taken for $B$ to execute (latency of $B$).
The latency of the \orchnet\ $\Orchnet=(\net,\Phi,T,\Tinit)$, for a given $\daemon$ 
is
\[
E_\daemon(\Orchnet) = E_\daemon(\maxconfig(\Orchnet,\daemon),\Orchnet)
\]
%
\section{Characterizing monotony}
\label{sec:char-monotony}

In this article, we are interested in the total time taken to 
execute a web-service orchestration. As a consequence, we will consider only
orchestrations that terminate in a finite time, \ie, only a finite
number of values can be returned.

\subsection{Defining monotony}

To formalize monotony we must specify how latencies and initial dates
can vary. As an example, we may want to constrain some pair of
transitions to have identical latencies, or some pair of minimal
places to have identical initial dates. This allowed flexibility in
setting latencies or initial dates is formalized under the notion of
\preorchnet\ we introduce next.

\begin{definition}[\preorchnet] \label {perifueopruif}
Call \emph{\preorchnet} a tuple
$\Preorchnet=(\net,\Phi,\datefamilies,\dateinit)$, where $\net$ and
$\Phi$ are as before, and $\datefamilies$ and $\dateinit$ are sets of
families $T$ of latency functions and of families $\Tinit$ of initial
date functions. Write $\Orchnet\in\Preorchnet$ if
$\Orchnet=(\net,\Phi,T,\Tinit)$ for some $T\in\datefamilies$ 
and $\Tinit\in\dateinit$.
\end{definition}
For two families $T$ and $T'$ of
latency functions, write
\[
T \geq T'
\]
to mean that $\forall \daemon\in\Daemons, \forall t\in\transset \implies
\tau_{t}(\daemon) \geq \tau'_{t}(\daemon) $, and similarly for 
$\Tinit\geq\Tinit'$. For $\Orchnet,\Orchnet'\in\Preorchnet$, write
\[
\Orchnet\geq\Orchnet'
\] 
if $T \geq T'$ and $\Tinit\geq\Tinit'$ both hold.

\begin{definition} [monotony] \label {0p3498ry398}
{\preorchnet} \mbox{$\Preorchnet=(\net,\Phi,\datefamilies,\dateinit)$}
is called
\emph{monotonic} if, for any two $\Orchnet,\Orchnet'\in\Preorchnet$, 
such that $\Orchnet\geq\Orchnet'
$, we have $E_\daemon(\Orchnet)\geq E_\daemon(\Orchnet')$.
\end{definition}
Considering legal sets $\datefamilies$ and $\dateinit$ of families of
latency functions and initial date functions allows setting
constraints on these families. This possibility will be essential in
characterizing monotonic orchestrations.

\subsection{A global necessary and sufficient condition}

\begin{theorem} \label {eropiufheluih} \
\begin{enumerate}
\item \label {er;ofuihel;uiof} The following condition implies the
monotony of \preorchnet\ $\Preorchnet=(\net,\Phi,\datefamilies,\dateinit)$:
%
\beq
\bea{ll}
\forall \Orchnet\in\Preorchnet,
&\forall\daemon\in\Daemons, \forall \maxconfig\in\MaxConfigs{\net},\\
&E_\daemon(\maxconfig,\Orchnet) 
\geq
E_\daemon(\maxconfig(\Orchnet,\daemon),\Orchnet) 
\eea
\label {rpeoifuerifh}
\eeq
where $\MaxConfigs{\net}$ denotes the set of all maximal
configurations of net $\net$ and $\maxconfig(\Orchnet,\daemon)$ is the
maximal configuration of $\Orchnet$ that actually occurs 
under the daemon value $\daemon$.

\item \label {periufgehl} Conversely, assume that: 
\begin{enumerate}
\item Condition
(\ref{rpeoifuerifh}) is violated, and 
\item\label{subcond:124mn2t23} for any two \orchnet s
$\Orchnet$ and $\Orchnet'$ such that $\Orchnet\in\Preorchnet$, then 
$\Orchnet'\geq\Orchnet\implies\Orchnet'\in\Preorchnet$. 
\end{enumerate}
Then
$\Preorchnet=(\net,\Phi,\datefamilies,\dateinit)$ is not
monotonic.
\end{enumerate}
\end{theorem}
Sub-condition b) in statement \ref{periufgehl} destroys the necessity
of Condition (\ref{rpeoifuerifh}). Statement \ref{periufgehl}
expresses that Condition (\ref{rpeoifuerifh}) is also necessary
provided that it is legal to increase at will latencies or initial
dates. Observe that Condition (\ref{rpeoifuerifh}) by itself cannot be
enough since it trivially holds if $\datefamilies$ is a singleton.

\proof We first prove Statement \ref{er;ofuihel;uiof}.
Let $\Orchnet'\in\Preorchnet$ be such that $\Orchnet'\geq\Orchnet$. We have:
\beqq
E_\daemon(\maxconfig(\Orchnet',\daemon),\Orchnet') &\geq&
E_\daemon(\maxconfig(\Orchnet',\daemon),\Orchnet) 
\\
&\geq&
E_\daemon(\maxconfig(\Orchnet,\daemon),\Orchnet) 
\eeqq
where the first inequality follows from the fact that
$\maxconfig(\Orchnet',\daemon)$ is a conflict free partial order and
$\Orchnet'\geq\Orchnet$, and the second inequality follows from
(\ref{rpeoifuerifh}) applied with
$\maxconfig=\maxconfig(\Orchnet',\daemon)$. This proves Statement
\ref{er;ofuihel;uiof}.

We prove statement \ref{periufgehl} by contradiction. Let
$(\Orchnet,\daemon,\maxconfig^\dagger)$ be a triple violating Condition
(\ref{rpeoifuerifh}), in that 
\beq
&&\maxconfig^\dagger \mbox{ cannot occur} 
\label {oerfuihe}
\\
&& E_\daemon(\maxconfig^\dagger,\Orchnet) 
<
E_\daemon(\maxconfig(\Orchnet,\daemon),\Orchnet) 
\label {repfuihpeuif}
\eeq
Now consider the \orchnet\ net $\Orchnet'=(\net,\Phi,T',\Tinit)$
where the family $T'$ is the same as $T$ except that in $\daemon$,
$\forall t\notin\maxconfig^\dagger$, 
$\tau'_t(\daemon) > E_\daemon(\maxconfig^\dagger,\Orchnet)$.
Clearly $\Orchnet'\geq\Orchnet$.
But using construction (\ref{eroifuego}), it is easy to verify
that $\maxconfig(\Orchnet',\daemon)=\maxconfig^\dagger$ and thus 
\beqq
E_\daemon(\maxconfig(\Orchnet',\daemon),\Orchnet') 
&=&
E_\daemon(\maxconfig^\dagger,\Orchnet')
\\
&=&
E_\daemon(\maxconfig^\dagger,\Orchnet)
\\
&<&
E_\daemon(\maxconfig(\Orchnet,\daemon),\Orchnet),
\eeqq
which violates monotony.
\eproof

\subsection{A structural condition for the monotony of 
workflow nets}
\newcommand{\morphism}{\chi}

\emph{Workflow nets}~\cite{Aalst97} were proposed as a simple model for
workflows. These are Petri nets, with a special minimal place $i$ and
a special maximal place $o$. We consider the class of workflow nets that
are 1-safe and which have no loops. Further, we require them to be 
\emph{sound}~\cite{Aalst97}. 

\begin{definition} A Workflow net $\wnet$ is said to be \emph{sound}
iff:
\begin{enumerate}
\item For every marking $M$ reachable from the initial place $i$,
there is a firing sequence leading to the final place $o$.
\item If a marking $M$ marks the final place $o$, then no other
place can in $\wnet$ can be marked in $M$
\item There are no \emph{dead} transitions in $\wnet$. Starting from
the initial place, it is always possible to fire any transition of
$\wnet$.
\end{enumerate}
\end{definition}
An example of workflow net is shown in the first net of
Figure~\ref{fig:monotone-eg}.
Workflow nets will be generically denoted by $\wnet$. We can equip
workflow nets with same attributes as occurrence nets, this yields
\emph{\prewfnet s}
$\PreWFnet=(\wnet,\Phi,\datefamilies,\dateinit)$. Referring to the end
of Section~\ref{perofuiheriofuh}, unfolding $\wnet$ yields an
occurrence net that we denote by $\net_\wnet$ with associated morphism
$\varphi_\wnet:\net_\wnet\mapsto\wnet$, see
Figure~\ref{fig:monotone-eg}. Here the morphism $\varphi_\wnet$ maps the
two $c$ transitions (and the place in its preset and postset) in the net on the right
to the single $c$ transition (and its preset and postset) in the net on the left.
Observe that $\wnet$ and $\net_\wnet$
possess identical sets of minimal places.  Morphism $\varphi_\wnet$
induces a \preorchnet\
\[
\Preorchnet_\wnet=(\net_\wnet,\Phi_\wnet,\datefamilies_\wnet,\dateinit)
\]
by attaching to each transition $t$ of $\net_\wnet$ the value and latency
functions attached to $\varphi_\wnet(t)$ in $\PreWFnet$.

We shall use the results of the previous section in order to
characterize those \prewfnet s whose unfoldings give monotonic \preorchnet s. Our
characterization will be essentially structural in that it does not
involve any constraint on latency functions beyond equality
constraints, for some pairs of transitions. Under
this restricted discipline, the simple structural conditions we shall
formulate will also be almost necessary.

We first define a notion of \emph{cluster} on safe nets, which will be
useful for our characterisation.
\begin{definition}[clusters] \label {fiuehfiu}
For a safe net $N$, a \emph{cluster} is a minimal set $\cluster$ of
places and transitions of $N$ such that
\beq
\forall t\in\cluster\implies\preset{t}\subseteq\cluster
&,& \forall p\in\cluster \implies \postset{p}\subseteq\cluster
\label {oruyfyegoruy}
\eeq
\end{definition}

\begin{theorem}[Sufficient Condition] \label {ptfgheiufh}
Let $\wnet$ be a \wfnet\ and $\net_\wnet$ be its unfolding.
A sufficient condition for the \preorchnet\ 
$\Preorchnet_\wnet=(\net_\wnet,\Phi_\wnet,\datefamilies_\wnet,\dateinit)$
to be monotonic is that every cluster $\cluster$ satisfies the
following condition:
\beq
\label {poiuh}
\forall t_1,t_2\in\cluster,\ t_1\ne t_2\implies \postset{t_1}=\postset{t_2}
\eeq
\end{theorem}
\proof Let $\varphi_\wnet$ be the net morphism mapping $\net_\wnet$
onto $\wnet$ and let $\Orchnet\in\Preorchnet$ be any \orchnet.  We
prove that condition~\ref{er;ofuihel;uiof} of
Theorem~\ref{eropiufheluih} holds for $\Orchnet$ by induction on the
number of transitions in the maximal configuration
$\maxconfig(\Orchnet,\daemon)$ that actually occurs. The base case is
when it has only one transition.  Clearly this transition has the
least latency and any other maximal configuration has a greater
execution time.

\paragraph*{Induction Hypothesis}
Condition~\ref{er;ofuihel;uiof} of Theorem~\ref{eropiufheluih}
holds for any maximal occurring configuration with $m-1$ transitions ($m > 1$). Formally,
for a \preorchnet\ $\Preorchnet=(\net,\Phi,\datefamilies,\dateinit)$:
$\forall \Orchnet\in\Preorchnet,
\forall\daemon\in\Daemons,
\forall \maxconfig\in\MaxConfigs{\net}$,
\beq
E_\daemon(\maxconfig,\Orchnet) 
\geq
E_\daemon(\maxconfig(\Orchnet,\daemon),\Orchnet) 
\label {rpeoifuerifh123r}
\eeq
holds if $|\{t \in \maxconfig(\Orchnet,\daemon)\}|\leq m-1$.

\paragraph*{Induction Argument}
Consider the \orchnet\ $\Orchnet$, where the actually occurring
configuration $\maxconfig(\Orchnet,\daemon)$ has $m$
transitions. $\config'$ is any other maximal configuration of
$\Orchnet$.  If the transition $t$ in $\maxconfig(\Orchnet,\daemon)$
with minimal date $d_t$ also occurs in $\config'$ then comparing
execution times of $\maxconfig(\Orchnet,\daemon)$ and $\config'$
reduces to comparing $E_{\daemon}(\maxconfig(\Orchnet,\daemon)
\setminus \{t\},\Orchnet^t)$ and $E_{\daemon}(\config' \setminus
\{t\}, \Orchnet^t)$.  Since $\maxconfig(\Orchnet,\daemon) \setminus
\{t\}$ is the actually occurring configuration in the future
$\Orchnet^t$ of transition $t$, using our induction hypothesis, we
have
\[ 
E_{\daemon}(\maxconfig(\Orchnet,\daemon) \setminus \{t\},\Orchnet^t)
\leq
E_{\daemon}(\config' \setminus \{t\}, \Orchnet^t)
\]
and so
\[
E_{\daemon}(\maxconfig(\Orchnet,\daemon),\Orchnet)\ \leq 
E_{\daemon}(\config',\Orchnet)
\]
If $t\notin\config'$ for some $\config'$, then there must exist another
transition $t'$ such that $\preset{t} \cap \preset{t'} \neq \emptyset$. 
By the definition of clusters, $\varphi_\wnet(t)$ and $\varphi_\wnet(t')$
must belong to the same cluster $\cluster$. Hence, 
$\postset{t}=\postset{t'}$ follows from
condition \ref{poiuh} of Theorem \ref{ptfgheiufh}. The futures
$\Orchnet^t$ and $\Orchnet^{t'}$ thus have identical sets of transitions:
they only differ in the initial marking of their places.
If $T_{init}$ and $T'_{init}$ are the initial marking of these places,
$T_{init} \leq T'_{init}$ (since $d_t \leq d_{t'}$, $\postset{t}$
has dates lesser than $\postset{t'}$).
Hence
\beq
E_{\daemon}(\maxconfig(\Orchnet,\daemon),\Orchnet)\ &=& 
E_{\daemon}(\maxconfig(\Orchnet,\daemon) \setminus \{t\},\Orchnet^t)
\label{eqn:r2f2t}
\eeq
and 
\beq
E_{\daemon}(\config',\Orchnet)\ &=& 
E_{\daemon}(\config' \setminus \{t'\},\Orchnet^{t'}) \nonumber \\
&\geq& E_{\daemon}(\config' \setminus \{t'\},\Orchnet^{t})
\label{eqn:r2f2tr23t}
\eeq
The inequality holds since $\Orchnet^{t'} \geq \Orchnet^{t}$.
The induction hypothesis on (\ref{eqn:r2f2t}) and (\ref{eqn:r2f2tr23t})
gives 
\[
E_{\daemon}(\maxconfig(\Orchnet,\daemon),\Orchnet)\ \leq 
E_{\daemon}(\config',\Orchnet)
\]
This proves the theorem.\eproof \\

\paragraph*{Comments about tightness of the conditions of
Theorem~\protect\ref{ptfgheiufh}} Recall that the sufficient condition
for monotony stated in Theorem~\ref{eropiufheluih} is ``almost
necessary'' in that, if enough flexibility exist in setting latencies
and initial dates, then it is actually necessary. It turns out that
the same holds for the sufficient condition stated in
Theorem~\protect\ref{ptfgheiufh} if the workflow net is assumed to be live.
%
%
\begin{theorem}[Necessary Condition]\label {erpifheif}
Suppose that the workflow net $\wnet$ is sound.
Assume that $\WFnet\in\PreWFnet$ and
$\WFnet'\geq\WFnet$ implies $\WFnet'\in\PreWFnet$, meaning that there
is enough flexibility in setting latencies and initial dates. In addition,
assume that there is atleast one $\WFnet^*\in\PreWFnet$ such that there is an
daemon value $\daemon^*$ for which the latencies of all the transitions are finite. Then the
sufficient condition of Theorem~\protect\ref{ptfgheiufh} is also
necessary for monotony.
\end{theorem}
\proof
\newcommand{\stcone}[1]{[#1]}
We will show that when condition (\ref{poiuh}) of Theorem
\ref{ptfgheiufh} is not satisfied by $\wnet$, the Orchnets in its
induced preOrchNet $\Preorchnet_\wnet$ can violate condition
(\ref{rpeoifuerifh}) of Theorem \ref{eropiufheluih}, the necessary
condition for monotony.

Let $c_\wnet$ be any cluster in $\wnet$ that violates the condition
\ref{poiuh} of Theorem \ref{ptfgheiufh}.  Consider the unfolding of
$\wnet$, $\net_\wnet$ and the associated morphism
$\varphi:\net_\wnet\mapsto\wnet$ as introduced before.  Since $\wnet$
is sound, all transitions in $c_\wnet$ are reachable from the initial
place $i$ and so there is a cluster $c$ in $\net_\wnet$ such that
$\varphi(c)=c_\wnet$. There are transitions $t_1, t_2 \in c$ such that
$\preset{t_1}\cap\preset{t_2}\ne\emptyset$,
$\preset{\varphi(t_1)}\cap\preset{\varphi(t_2)}\ne\emptyset$ and
$\postset{\varphi(t_1)}\ne\postset{\varphi(t_2)}$.  Call
$\stcone{t}=\cone{t}\setminus\{t\}$ and define
$K=\stcone{t_1}\cup\stcone{t_2}$. We consider the following two cases:

\paragraph{$K$ is a configuration}

If so, consider the OrchNet $\Orchnet^*\in\Preorchnet_\wnet$ got when
transitions of $\net_\wnet$ (and so $\wnet$) have latencies as that in
$\WFnet^*$.  So for the daemon value $\daemon^*$, the quantity
$E_{\daemon^*}(K,\Orchnet^*)$ is some finite value $\mathbf{n}^*$.  Now, configuration $K$ can actually occur in a OrchNet
$\Orchnet$, such that $\Orchnet > \Orchnet^*$, where $\Orchnet$ is obtained as
follows ($\tau$ and $\tau^*$ denote the latencies of transitions in
$\Orchnet$ and $\Orchnet^*$ respectively): $\forall t\in K, t'\in\net_\wnet$
s.t.  $\preset{t}\cap\preset{t'}\ne\emptyset$, set $\tau_{t'}(\daemon^*)=
\mathbf{n}^*+1$ and keep the other latencies unchanged. In this case, for the
daemon value $\daemon^*$, the latencies of  all transitions of $\Orchnet$ (and
so its overall execution time) is finite.
Denote by $\Orchnet^K$ the future of $\Orchnet$ once  configuration $K$ has
actually occurred.  Both $t_1$ and $t_2$ are minimal and enabled in $\Orchnet^K$.

 Since $\postset{\varphi(t_1)}\ne\postset{\varphi(t_2)}$, without loss of generality, we assume
that there is a place $p\in\postset{t_1}$ such that  $\varphi(p)\in\postset{\varphi(t_1)}$ but
$\varphi(p)\notin\postset{\varphi(t_2)}$. Let $t^*$ be a transition in
$\Orchnet^K$ such that $t^*\in\postset{p}$. Such a transition must
exist since $p$ can not be a maximal place: $\varphi(p)$ can not be a
maximal place in $\wnet$ 
which has a unique maximal place. Now consider the Orchnet
$\Orchnet'>\Orchnet$ got as follows:
$\tau'_{t_1}(\daemon^*)=\tau_{t_1}(\daemon^*), \tau'_{t_2}(\daemon^*) = \tau_{t_1}(\daemon^*)+1$ 
and for all other $t\in c, \tau'_t(\daemon^*) = \tau'_{t_2}(\daemon^*)+1$.  Set
$\tau'_{t^*}(\daemon^*)=\infty$ and for all other transitions of $\Orchnet'$,
the delays are the same as that in $\Orchnet$ and thus are finite for $\daemon^*$.

 $t_1$ has the minimal delay among all transitions in $c$, 
and $t^*$ is in the future of $t_1$. So the actually occuring
configuration $E_{\daemon^*}(\maxconfig(\Orchnet',\daemon^*),\Orchnet')$ has an infinite delay.
However any maximal configuration $\maxconfig$ which does not include
$t_1$ (for eg, when $t_2$ fires instead of $t_1$) will have a finite
delay.  For such $\maxconfig$ we thus
have $E_{\daemon^*}(\maxconfig(\Orchnet',\daemon^*),\Orchnet') >
E_{\daemon^*}(\maxconfig,\Orchnet')$ and so $\Orchnet'$ violates the
condition (\ref{rpeoifuerifh}) of Theorem \ref{eropiufheluih}.

\paragraph{$K$ is not a configuration}
If so, there exist transitions $t\in\stcone{t_1}\setminus\stcone{t_2},\
t'\in\stcone{t_2}\setminus\stcone{t_1}$ such that $\preset{t}\cap
\preset{t'}\ne\emptyset$, $\preset{\varphi(t)}\cap
\preset{\varphi(t')}\ne\emptyset$ and
$\postset{\varphi(t)}\ne\postset{\varphi(t')}$. The final condition holds
since $t_2$ and $t_1$ are not in the causal future of $t$ and $t'$
respectively.
Thus $t$ and $t'$ belong to the same cluster, which violates condition
\ref{poiuh} of Theorem \ref{ptfgheiufh} and we can apply the same reasoning as
in the beginning of the proof. Since $\stcone{t}$ is finite for any transition
$t$, we will eventually end up with $K$ being a configuration.  \eproof

\subsection{Discussion regarding monotony}
Based on the mathematical results of the former section, we shall
first discuss which orchestrations are monotonic. The resulting class
is not very wide, as we shall see. However, further thinking suggests
that considering QoS parameters in isolation --- as we did so far ---
is not the right way to proceed. We will thus revisit QoS and
monotony.

\paragraph{Which orchestrations are monotonic? \color{white}}
Not surprisingly, orchestrations involving no choice at all --- they
are modeled by event graphs --- are monotonic. 

Theorem \ref{ptfgheiufh}, however, allows for more general
orchestrations. In particular, if multi-threading with only conflicting
threads are used, then the theorem essentially requires that the
choice among the different threads is performed, based on their
overall latency (where a thread is considered as atomic). It is
allowed, however, that conflicting threads possess different overall
latencies. Next, if a blend of concurrent and conflicting
multi-threading is used, then 1/ concurrent threads should have equal
latencies, and 2/ the choice among conflicting threads should be based
on their overall latency. Finally, concurrent multi-threading (with no
conflict with other thread when the concurrent threads are launched)
raises no problem. An example of a non trivial monotonic orchestration
is shown in Figure \ref{fig:monotone-eg}.

\paragraph{Revisiting QoS and monotony}
As discussed before, orchestrations involving data dependent control
will very often be non monotonic. This makes SLA/SLS/contract based
QoS management very problematic. What did we wrong?

In fact, the problem is that we consider QoS parameters in isolation
when dealing with SLS or contracts. However 
one can trade latency for ``quality of data''. 
Typically, by reducing timeouts while waiting for responses
from called Web services, one can easily reduce latency. But there is
no free lunch: if one does so, then exceptions get raised in lieu of
getting valid responses for the query.

So we must refine our notion of monotony as follows: say that an
orchestration is ``conditionally monotonic'' if, \emph{under the
constraint that identical responses are received by the
orchestration,} increasing some latency will cause an increase of the
overall latency of the orchestration. In the next section we formalize
this notion and study the characterization of conditional monotony.

\section{Refined QoS and Conditional monotony}
\label{sec:refined-qos}

Define the set of values returned by a maximal configuration $\config\in\MaxConfigs{\Orchnet}$ as 
\[
V_{\daemon}(\config,\Orchnet)\ =\ \{\phi_t(\daemon)\ |\ t\in\mathrm{max}(\config)\}
\]
where $\mathrm{max}(\config)$ denotes the maximal (w.r.t $\leq$)
transitions in the configuration $\config$.
The values returned by the orchestration $\Orchnet=(\net,\phi,T,\Tinit)$ for a given
value of the daemon $\daemon$ is
\[
V_{\daemon}(\Orchnet)\ =\ V_{\daemon}(\maxconfig(\Orchnet,\daemon),\Orchnet)
\]

\begin{definition} [conditional monotony] \label {;ewfouihw;oif}
{\preorchnet} \mbox{$\Preorchnet=(\net,\Phi,\datefamilies,\dateinit)$}
is called \emph{conditionally monotonic} if, 
\beq
\left.
\bea{l}
\forall \Orchnet,\Orchnet'\in\Preorchnet \mathrm{\ s.\ t.\ }
\Orchnet \geq \Orchnet' \mathrm{\ and}
\\
\forall \daemon\in\Omega\ \mathrm{\ s.\ t.\ }
V_{\daemon}(\Orchnet)=V_{\daemon}(\Orchnet')
\eea
\right\}
\Rightarrow E_\daemon(\Orchnet)\geq E_\daemon(\Orchnet')
\nonumber
\eeq
\end{definition}
%
%
Definition \ref{;ewfouihw;oif} does not attempt to compare overall
orchestration latencies, if different responses are received as a
result of changing latencies of called Web services. In particular,
cases where valid data are received are not compared with cases where
exceptions are raised. Said differently, Definition \ref{;ewfouihw;oif}
prohibits trading latency for quality of data.

Conditional monotony does not seem to be considered while formulating WSLA contracts.
The 'conditional' contracts in WSLA are not related to the notion of
conditional monotony but rather refer to the obligations that need
to be met by the client for the server's promises to be fulfilled.

\begin{assumption} \label {pweifuhwpiuf}
We assume that \orchnet\ $\net$ is such that for all distinct $\config,\config' \in\MaxConfigs{\Orchnet}$,
$V_{\daemon}(\config,\Orchnet)\ne V_{\daemon}(\config',\Orchnet)$.
\end{assumption}
Assumption \ref{pweifuhwpiuf} is very natural when normal processing
of the orchestration together with cases where exceptions occur are
considered. With this assumption, Theorem \ref{eropiufheluih} 
simplifies as follows:
\begin{theorem} \label {repfiouehpuio}
Under Assumption \ref{pweifuhwpiuf} \preorchnet\
$\Preorchnet=(\net,\Phi,\datefamilies,\dateinit)$ is conditionally monotonic.
\end{theorem}
\proof
In fact this is a trivial result: by Assumption \ref{pweifuhwpiuf}, we
do not need to compare latencies across different configurations.  But
each configuration possesses no conflict and is therefore an event
graph. Since dating in event graphs is amenable of $\max/+$ algebra,
it is monotonic. Thus \preorchnet\
$\Preorchnet=(\net,\Phi,\datefamilies,\dateinit)$ is conditionally monotonic.
\eproof

\section{Conclusion}
\label{sec:conc}
This paper is a contribution to the fundamentals of contract based QoS
management of Web services orchestrations/choreographies. QoS
contracts implicitly assume monotony w.r.t. QoS parameters. This paper
focuses on one specific (but representative) QoS parameter, namely
latency or response time.  We have shown that monotony is very easily
violated in realistic cases. We have formalized monotony and have
provided necessary and sufficient conditions for it. As we have seen,
QoS can be very often traded for Quality of Data: poor quality
responses to queries (including exceptions or invalid responses) can
be obtained typically much faster. This reveals that QoS parameters
should not be considered separately, in isolation. We have thus
revisited the notions of latency and monotony and proposed the new
concept of \emph{conditional monotony.} Fortunately enough, typical
orchestrations turn out to be non-monotonic, but conditionally
monotonic.

Our study has an impact on the way SLS should be phrased for
orchestrations: it suggests crude contracts such as ``{for 95\% of the
requests the response time will be less than 5ms}'' should not be used
without referring to the quality of data. 
Also, our
study has an impact on contract monitoring --- monitoring whether a
called service meets its contract.

We see two extensions of this work. 
First, our mathematical results rely on the notion of branching cells,
which were developed for nets without read arcs. However, advanced
orchestration languages such as Orc~\cite{orc} offer a sophisticated
form of preemption that requires contextual nets (with read arcs) for
their modeling. Extending our results to this case is non trivial
since branching cells do not exist for contextual nets.
Second, this paper considers latencies in a non probabilistic context.
However, these authors advocated in~\cite{QoS-icws07} the use of
probabilistic contracts --- they better reflect the uncertain nature
of QoS parameters in the open world of the Web and they allow for well
sound overbooking and less pessimistic contracts. We should extend our
present work to this probabilistic framework. This should not be too
difficult, as we guess. In fact, a theory of monotony for
probabilistic QoS must rely on an order among probability
distributions. This exists and is known as \emph{stochastic ordering.}
Results are available that relate stochastic ordering and point-wise
ordering (as we study here), which should be of help for such an
extension.

\bibliographystyle{plain}
\bibliography{references}

\begin{thebibliography}{10}

\bibitem{aalst98application}
{W.M.P. van der} Aalst.
\newblock {The Application of Petri Nets to Workflow Management}.
\newblock {\em The Journal of Circuits, Systems and Computers}, 8(1):21--66,
  1998.

\bibitem{bpel4wsspec}
Tony Andrews, Francisco Curbera, Hitesh Dholakia, Yaron Goland, Johannes Klein,
  Frank Leymann, Kevin Liu, Dieter Roller, Doug Smith, Satish Thatte, Ivana
  Trickovic, Sanjiva Weerawarana, and Satish Thatte.
\newblock {Business Process Execution Language for Web Services Version 1.1}.
\newblock Specification, BEA Systems, International Business Machines
  Corporation, Microsoft Corporation, SAP AG, Siebel Systems, May 2003.

\bibitem{Fisteus2005}
Jes{\'u}s Arias-Fisteus, Luis~S{\'a}nchez Fern{\'a}ndez, and Carlos~Delgado
  Kloos.
\newblock {Applying model checking to BPEL4WS business collaborations}.
\newblock In {\em SAC}, pages 826--830, 2005.

\bibitem{WSLA}
Alexander Keller and Heiko Ludwig.
\newblock The wsla framework: Specifying and monitoring service level
  agreements for web services.
\newblock {\em J. Network Syst. Manage.}, 11(1), 2003.

\bibitem{orctrace}
David Kitchin, William~R. Cook, and Jayadev Misra.
\newblock A language for task orchestration and its semantic properties.
\newblock In {\em CONCUR}, pages 477--491, 2006.

\bibitem{MarsanBCDF98}
Marco~Ajmone Marsan, Gianfranco Balbo, Gianni Conte, Susanna Donatelli, and
  Giuliana Franceschinis.
\newblock Modelling with generalized stochastic petri nets.
\newblock {\em SIGMETRICS Performance Evaluation Review}, 26(2):2, 1998.

\bibitem{orc}
Jayadev Misra and William~R. Cook.
\newblock Computation orchestration: A basis for wide-area computing.
\newblock {\em Journal of Software and Systems Modeling}, May, 2006.
\newblock Available for download at {\tt
  http://dx.doi.org/10.1007/s10270-006-0012-1}.

\bibitem{bpmBpelAnalysisReport05}
C.~Ouyang, E.~Verbeek, W.M.P van~der Aalst, and S.~Breutel.
\newblock Formal {S}emantics and {A}nalysis of {C}ontrol {F}low in {WS-BPEL}.
\newblock BPM Center Report BPM-05-15, BPMcenter.org, 2005.

\bibitem{QoS-icws07}
Sidney Rosario, Albert Benveniste, Stefan Haar, and Claude Jard.
\newblock Probabilistic {Q}o{S} and soft contracts for transaction based web
  services.
\newblock In {\em ICWS}, pages 126--133, 2007.

\bibitem{orc-wsfm07}
Sidney Rosario, David Kitchin, Albert Benveniste, William Cook, Stefan Haar,
  and Claude Jard.
\newblock Event structure semantics of orc.
\newblock In {\em WSFM}, 2007.

\bibitem{Aalst97}
Wil M.~P. van~der Aalst.
\newblock Verification of workflow nets.
\newblock In {\em ICATPN}, pages 407--426, 1997.

\end{thebibliography}


\end{document}